\documentstyle[sprocl,psfig]{article}

\def\Journal#1#2#3#4{{#1} {\bf #2}, #3 (#4)}

\def\NPB{{\em Nucl. Phys.} B}
\def\PLB{{\em Phys. Lett.}  B}
\def\PRL{\em Phys. Rev. Lett.}
\def\PRD{{\em Phys. Rev.} D}
\def\ZPC{{\em Z. Phys.} C}
\def\PTP{\em Prog. Theor. Phys.}

\def\be{\begin{equation}}
\def\ee{\end{equation}}
\def\bea{\begin{eqnarray}}
\def\eea{\end{eqnarray}}

\newcommand{\ol}[1]{\overline{#1}}

\newcommand{\tanb}{\tan\beta}
\newcommand{\bb}{$B^0$--$\ol{B}^0$}
\newcommand{\kk}{$K^0$--$\ol{K}^0$}
\newcommand{\ek}{\epsilon_K}
\newcommand{\bsg}{b\rightarrow s\, \gamma}

\newcommand{\klpnn}{K_L\rightarrow \pi^0\, \nu\, \ol{\nu}}
\newcommand{\kppnn}{K^+\rightarrow \pi^+\, \nu\, \ol{\nu}}
\newcommand{\dmbd}{\Delta M_{B_d}}
\newcommand{\dmbs}{\Delta M_{B_s}}

\newcommand{\bsbs}{$B_s$--$\bar{B}_s$}
\newcommand{\bdbd}{$B_d$--$\bar{B}_d$}
\newcommand{\meg}{\mu \to e\,\gamma}
\newcommand{\tmg}{\tau \to \mu\,\gamma}

\newcommand{\dmbsd}{\Delta m_s/\Delta m_d}

\begin{document}

\title{Manifestaion of SUSY in B decays
\footnote{
Talk given at the Third International Conference on B
Physics and CP Violation, December 3 -7, 1999, Taipei. 
}}

\author{YASUHIRO OKADA}

\address{Institute for Particle and Nuclear Studies, KEK,\\
Oho 1-1, Tsukuba 305-0801, Japan\\E-mail: yasuhiro.okada@kek.jp} 




\maketitle\abstracts{SUSY effects on various flavor
changing neutral current processes are discussed in
the minimal supergravity model and the SU(5) grand unified theory 
with right-handed neutrino supermultiplets.
In particular, in the latter case the neutrino Yukawa coupling 
constants can be a source of the flavor mixing in the 
right-handed-down-type-squark sector. It is shown that due to this
mixing the time-dependent CP asymmetry of radiative $B$ decay can be as
large as 30\% and the ratio of \bsbs\ mixing and \bdbd\ mixing 
deviates from the prediction in the standard model and the minimal
supergravity model without the neutrino interaction.
}

\section{Introduction}

In order to explore supersymetry (SUSY) indirect search experiments 
can play a complementary role to direct search for SUSY particles
at collider experiments.
Since SUSY particles may affect flavor changing neutral current
(FCNC) processes and CP violation in B and K meson decays,
it is possible that new experiments in B decay at both $e^+ e^-$
colliders and hadron machines reveal new physics signals which
can be interpreted as indirect evidence of SUSY.
 
In the context of SUSY models flavor physics has important 
implications. Because the squark and the slepton
mass matrices become new sources of flavor mixings 
generic mass matrices would induce too large FCNC 
and lepton flavor violation (LFV) effects if the superpartners' 
masses are in a few-hundred-GeV
region. For example, if we assume that the SUSY contribution to the  
$K^0 - \bar{K}^0$ mixing is suppressed because of the cancellation
among the squark contributions of different generations, the squarks
with the same gauge quantum numbers must
be highly degenerate in masses at least for the first two generations.

There are several scenarios to solve this flavor problem.
In the minimal supergravity model flavor problem are avoided
by taking SUSY soft-breaking terms as flavor-blind structure. 
The scalar mass terms are assumed to be common for
all scalar fields at the Planck  scale and therefore there are no 
FCNC effects nor LFV from 
the squark and slepton sectors at this scale. 
The physical squark and slepton masses are determined taking account 
of renormalization effects from the Planck to the weak scale. 
This induces sizable SUSY contributions to various FCNC and LFV 
processes. 

In this talk we consider two types of SUSY models and 
discuss FCNC processes. The first one is 
the minimal supersymmetric standard model (MSSM) with a 
universal SUSY breaking terms at the Planck scale which is
realized in the minimal supergravity model. The other is
the SU(5) grand unified theory with right-handed neutrino 
supermultiplets. This model incorporates the see-saw mechanism 
for neutrino mass generation. In the latter case the neutrino 
Yukawa coupling constants can be a source of the flavor mixing
in the right-handed-down-type-squark sector and due to this mixing 
the time-dependent CP asymmetry of radiative $B$ decay can be as
large as 30\% and the ratio of \bsbs\ mixing and \bdbd\ mixing 
deviates from the prediction for the standard model (SM) and the 
MSSM without the neutrino interaction.

\section{Update of FCNC Processes in the Supergravity Model}
In the minimal SM various FCNC processes and CP violation in B and K decays
are determined by the Cabibbo-Kobayashi-Maskawa (CKM) matrix. Constraints
on the parameters of the CKM matrix can be conveniently expressed
in terms of the unitarity triangle. With CP violation at B factory as well as
rare K decay experiments we will be able to check consistency of the 
unitarity triangle and at the same time search for effects of physics 
beyond the SM. In order to distinguish possible new physics effects it 
is important to identify how various models can modify the SM predictions.
 
Although general SUSY models can change the lengths and the angles of the 
unitarity triangle  in variety ways, the supergravity model predicts a 
specific pattern of the deviation from the SM.\cite{FCNC} 
Namely, we can show that the SUSY loop contributions to FCNC amplitudes 
approximately have the same dependence on the CKM elements as the SM 
contributions. In particular, if we assume that there are no CP 
violating phases from SUSY breaking sectors, the complex phase of 
the $B^0 - \bar{B^0}$ mixing amplitude does not change even if we 
take into account the SUSY contributions. The case with supersymmetric 
CP phases was also 
studied within the minimal supergravity model and it was shown that 
effects of new CP phases on the $B^0 - \bar{B^0}$ mixing amplitude 
and the direct CP asymmetry in the $\bsg$ process are small once 
constraints from neutron and electron EDMs 
are included.\cite{Nihei} 

We calculate various FCNC processes in the supergravity
model with universal soft breaking terms at a high energy scale. 
The results are summarized as follows.
\begin{enumerate}
\item The amplitude for $b \rightarrow s\gamma$ can receive a large
contribution from the SUSY and the charged-Higgs-top-quark loop diagrams.
The experimental branching ratio puts a strong constraint on SUSY parameter 
space. Since the SUSY contribution can interfere with other contributions
either constructively or destructively we cannot exclude 
the light charged Higgs boson region unlike the non-SUSY type II
two Higgs doublet model.\cite{bsg}
\item When the sign of the $b \rightarrow s\gamma$ amplitude is opposite 
to that of the SM, B($b \rightarrow sl^+l^-$) can be twice larger than the
SM prediction. This can occur for a large $\tanb$ region where 
$\tanb$ is the ratio of two vacuum expectation values of Higgs fields.
The deviation is also evident in the differential branching ratio and
the lepton forward-backward asymmetry.\cite{bsll}
\item In terms of consistency check of the unitarity triangle
the supergravity model has the following features.\cite{bbbar} 
(i)$\dmbd$ and $\ek$ are enhanced by the SUSY and charged-Higgs loop 
effects. When these quantities are normalized by the corresponding 
quantities in the SM they are almost independent of the CKM matrix element, 
and the enhancement factors for $\dmbd$ and $\ek$ are almost equal.
(ii) The branching ratios for $\kppnn$ and $\klpnn$ processes are 
suppressed compared to the SM prediction.  Again the suppression factor are
almost the same for two branching ratios and does not depend strongly on 
the CKM matrix element.
(iii) CP asymmetries in various B decay modes such as $B \rightarrow
J/\psi K_S$ and the ratio of $\dmbs$ and $\dmbd$ are the same as 
the SM prediction. 
\end{enumerate}
\begin{figure}
\begin{center}
\mbox{\psfig{figure=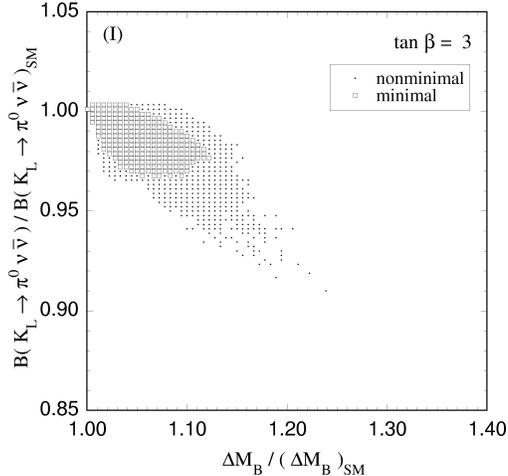,width=2.6in,angle=0}}
\end{center}
\caption{Correlation between $\Delta M_{B_d}$ and $\epsilon_K$ 
normalized by the SM value for $\tan{\beta}=3$.
The square(dot) points correspond to the minimal (enlarged) parameter space of
the supergravity model.
\label{fig:fig1}}
\end{figure} 
In Fig.~1 we present the correlation between $\dmbd$ and B($\klpnn$)
normalized by the corresponding quantities in the SM for $\tanb$ = 3.   
The constraint on the SUSY parameter space from the recent improved
SUSY Higgs search is implemented.\cite{bbbar} 
We have calculated the SUSY particle spectrum based on
two different assumptions on the initial conditions of renormalization 
group equations. 
The minimal case corresponds to the minimal supergravity 
where all scalar fields have a common SUSY breaking mass at the GUT scale. 
For ``nonminimal'' we enlarge 
the SUSY parameter space by relaxing the initial conditions for the 
SUSY breaking parameters, namely all squarks and sleptons have a 
common SUSY breaking mass whereas an independent
SUSY breaking parameter is assigned for Higgs fields.
The square(dot) points correspond to the minimal (enlarged) parameter 
space of the supergravity model. We can see that the  $\dmbd$ (and $\ek$) 
can deviated from the SM by 20\% whereas the deviation in $\klpnn$ and 
$\kppnn$ processes are small. 
These deviations may be evident in future when B factory experiments 
provide additional information on the CKM parameters. 

\section{FCNC in SUSY GUT with Right-handed Neutrino}
In this section we consider FCNC and LFV 
of charged lepton decays in
the model of a SU(5) SUSY GUT
which incorporates the see-saw mechanism for the neutrino mass
generation.\cite{baek}
In this model sources of the flavor mixing are Yukawa coupling
constant matrices for quarks and leptons as well as that for the
right-handed neutrinos.
Because the quark and lepton sectors are related by GUT interactions,
the flavor mixing relevant to the CKM matrix
can generate LFV such as $\meg$ and $\tmg$ processes \cite{LFV}
in addition to FCNC in hadronic observables.\cite{FCNC2} 
In the SUSY model with right-handed neutrinos, 
branching ratios of the LFV processes can become large enough to be measured 
in near-future experiments. \cite{nuR-LFV}
When we consider the right-handed neutrinos in the context of GUT, the
flavor mixing related to the neutrino oscillation can be a source of the
flavor mixing in the squark sector.
We show that due to the large mixing of the second and third generations
suggested by the atmospheric neutrino anomaly, \bsbs mixing, the
time-dependent CP asymmetry of the $B \rightarrow M_s\,\gamma$ process, where
$M_s$ is a CP eigenstate including the strange quark, can have a large
deviation from the SM prediction.\cite{baek}

The Yukawa coupling part and the Majorana mass term of the
superpotential for the SU(5) SUSY GUT with right-handed neutrino
supermultiplets is given by
$
  W =
    \frac{1}{8}f_U^{ij} \Psi_{i} \Psi_{j} H_{5}
  + f_D^{ij} \Psi_{i} \Phi_{j} H_{\bar{5}}
  + f_N^{ij} N_{i} \Phi_{j} H_{5}
  + \frac{1}{2} M_{\nu}^{ij} N_{i} N_{j}
$
,where $\Psi_{i}$, $\Phi_{i}$ and $N_{i}$ are ${\bf 10}$,
$\bar{\bf 5}$ and ${\bf 1}$ representations of SU(5) gauge group.
$i,j=1,2,3$ are the generation indices.
$H_{5}$ and $H_{\bar{5}}$ are Higgs superfields with ${\bf 5}$ and
$\bar{\bf 5}$ representations.

The renormalization effects due to the Yukawa
coupling constants induce various FCNC and LFV effects from the
mismatch between the quark/lepton and squark/slepton diagonalization
matrices.
In particular the large top Yukawa coupling constant is responsible for
the renormalization of the $\tilde{q}_L$ and $\tilde{u}_R$ mass
matrices.
At the same time the $\tilde{e}_R$ mass matrix receives sizable
corrections between the Planck and the GUT scales and various LFV
processes are induced.
In a similar way, if the neutrino Yukawa coupling constant $f_N^{ij}$ is
large enough, the $\tilde{l}_L$ mass matrix and the $\tilde{d}_R$ mass
matrix receive sizable flavor changing effects due to renormalization
between the Planck and the right-handed neutrino mass scales and the Planck 
and the GUT scales, respectively.
These are sources of extra contributions to LFV processes and various
FCNC processes such as $\bsg$, the \bb\ mixing and the \kk\ mixing.

The flavor mixing in the $\tilde{d}_R$ sector can induce large
time-dependent CP asymmetry in the $B \rightarrow M_s\,\gamma$ process.
Using the Wilson coefficients $c_7$ and $c'_7$ in the effective
Lagrangian for the $\bsg$ decay
$
{\cal L} = c_7 \bar{s} \sigma^{\mu\nu} b_R F_{\mu\nu}
          +c'_7 \bar{s} \sigma^{\mu\nu} b_L F_{\mu\nu} + H.c.
$,
the asymmetry is written as
$
  \frac{\Gamma(t) - \bar{\Gamma}(t)}{\Gamma(t) + \bar{\Gamma}(t)} =
  \xi A_t \sin\Delta m_d t
,
~~~
   A_t =
   \frac{2\rm{Im} ({\rm e}^{-i\theta_B} c_7 c'_7)}{|c_7|^2 + |c'_7|^2}
,
$
where $\Gamma(t)$ ($\bar{\Gamma}(t)$) is the decay width of
$B^0(t) \rightarrow M_s\,\gamma$ 
($\bar{B}^0(t) \rightarrow M_s\,\gamma$) and $M_s$ is some
CP eigenstate ($\xi = +1(-1)$ for a CP even (odd) state) such as
$K_S\, \pi^0$.\cite{AGS-CHH}
$\Delta m_d = 2|M_{12}(B_d)|$ and $\theta_B = \arg M_{12}(B_d)$ where
$M_{12}(B_d)$ is the \bdbd\ mixing amplitude.
Because the asymmetry can be only a few percent in the SM, a sizable
asymmetry is a clear signal of new physics beyond the SM.

We calculated various FCNC and LFV observables in this model under 
the assumption of the universal soft breaking terms at the Planck scale.
As typical examples of the neutrino parameters, we consider the
following parameter set corresponding to  the
Mikheyev-Smirnov-Wolfenstein (MSW) small mixing case.
$
  m_{\nu} =
  \left(
    2.236 \times 10^{-3},\,
    3.16 \times 10^{-3},\,
    5.92 \times 10^{-2}
  \right) ~{\rm eV}
$ and the Maki-Nakagawa-Sakata (MNS) matrix is given by
\begin{eqnarray}
  V_{\rm{MNS}} &=&
  \left(
    \begin{array}{ccc}
         0.999
      &  0.0371
      &  0
      \\ -0.0262
      &  0.707
      &  0.707
      \\ 0.0262
      &  -0.707
      &  0.707
    \end{array}
  \right)
~.
\nonumber
\end{eqnarray}
We also take $M_{\nu}$ to be proportional to a unit
matrix with a diagonal element of $M_R = 4 \times 10^{14}$ GeV.
We fix CKM parameters as $V_{cb}=0.04$, $|V_{ub}/V_{cb}|=0.08$ and
$\delta_{13}=60^{\circ}$.
We take $\tan\beta=5$ and vary other SUSY parameters.
We take account of various phenomenological constraints
on SUSY parameters including B($\bsg$).
We also calculated B($\meg$) and $\ek$ and imposed
constraints from these quantities.
\begin{figure}
\begin{center}
\mbox{\psfig{figure=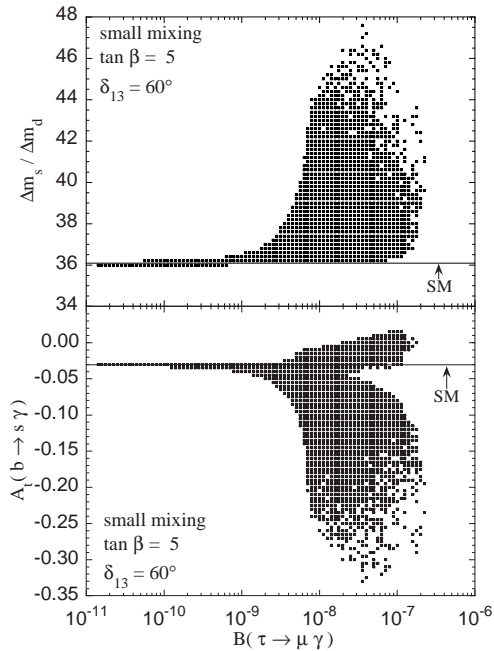,width=2.6in,angle=0}}
\end{center}
\caption{The ratio of \bsbs\ and \bdbd\ mass splittings $\dmbsd$ and
the magnitude factor $A_t$ of the time-dependent CP asymmetry in the
$B\rightarrow M_s\,\gamma$ process as a function of B($\tmg$)
for the small mixing.
\label{fig:fig2}}
\end{figure} 
The upper part of Fig.2 shows a correlation between
$\dmbsd$ (ratio of \bsbs\ and \bdbd\ mass splittings) and B($\tmg$).
We can see that $\dmbsd$
can be enhanced up to 30\% compared to the SM prediction.
This feature is quite different from the minimal supergravity model
without the GUT and right-handed neutrino interactions \cite{bbbar} where
$\dmbsd$ is almost the same as the SM value.
$A_t$ for the same parameter
set is shown as a function of B($\tmg$) in the lower part of
Fig. 2.
We can see that $|A_t|$ can be close to 30\% when B($\tmg$)is 
larger than $10^{-8}$.
The large asymmetry arises because the renormalization effect due to 
$f_N$ induces sizable contribution to $c'_7$ through
gluino--$\tilde{d}_R$ loop diagrams.
In this example possible new physics signals in B($\tmg$),
\bsbs\ mixing and $A_t$  all come from the renormalization effect 
on squark and slepton mass matrices from the large neutrino Yukawa coupling 
constant. Because these signals provide quite different signatures 
compared to the SM and the minimal supergravity model without GUT and 
right-handed neutrino interactions, future experiments in $B$ physics 
and LFV can provide us important clues on the interactions at very high
energy scale.

The work was supported in part by the Grant-in-Aid of the Ministry 
of Education, Science, Sports and Culture, Government of Japan (No.09640381),
Priority area ``Supersymmetry and Unified Theory of Elementary Particles'' 
(No. 707), and ``Physics of CP Violation'' (No.09246105).

\end{document}